\documentclass{revtex4}
\usepackage {amssymb}
\usepackage {amsmath}
\usepackage {epsfig}
\usepackage{graphicx}
\usepackage{hyperref}

\def\be {\begin{equation}}
\def\ee {\end{equation}}
\def\beq{\begin{eqnarray}}
\def\eeq{\end{eqnarray}}
\def\bea{\begin{eqnarray}}
\def\eea{\end{eqnarray}}
\def\nonu{\nonumber}
\begin{document}

\markboth{M. Horta\c{c}su, B.C. L\"{u}tf\"{u}o\u{g}lu and F.
Ta\c{s}k\i n}{Gauged System Mimicking the G\"{u}rsey Model}

\title{Gauged System Mimicking the G\"{u}rsey Model}

\author{M. Horta\c{c}su $^{1,2}$}
\email{hortacsu@itu.edu.tr}%

\author{B.C. L\"{u}tf\"{u}o\u{g}lu$^{1}$}
\email{bcan@itu.edu.tr}%

\author{F. Ta\c{s}k\i n$^{1,3}$}
\email{taskinf@itu.edu.tr}%

\affiliation{$^1$ Department of Physics, Istanbul Technical
University, Istanbul, Turkey,}

\affiliation{$^2$ Feza G\"{u}rsey Institute, Istanbul, Turkey}

\affiliation{$^3$ Department of Physics, Erciyes University, Kayseri, Turkey.}%

\date{\today}%

\begin{abstract}
We comment on the changes in the constrained model studied earlier
when constituent massless vector fields are introduced.  The new
model acts like a gauge-Higgs-Yukawa system, although its origin
is different.

\end{abstract}

\maketitle


\section{Introduction}

There are two endeavors which are studied by many physicists over
and over again, and often the solutions sought for the two
problems intersect.   The first one is to write a theory which is
shown to have a nontrivial limit, or  zero for the beta function
of the renormalization group for a non zero value of the coupling
constant, as the renormalization cut-off goes to infinity.   The
perturbatively nontrivial $\phi^4$ theory in four dimensions was
shown to go to a free theory as the cut-off was lifted a while ago
\cite{ba_ki_79,ba_ki_81,wi_73}.  This example shows that
perturbative expansions are not decisive in obtaining a nontrivial
model. There is continuing research on this subject \cite{kl_06}.

The other ongoing endeavor is to use only fermions to build a
model of nature, where all the observed bosons are constructed as
composites of these entities. In solid state physics the basic
fields, the electrons come together to form bosons to explain
superconductivity \cite{ba_co_sc_57}. Heisenberg spent years to
formulate a "theory of everything" for particle physics, using
only fermions \cite{he_54}.  The Nambu-Jona-Lasinio model
\cite{na_jo_61}, which was constructed based on an analogy with
the BCS theory of superconductivity \cite{ba_co_sc_57}, is a model
which satisfies both ambitions, which is written in terms of
fermions only and perturbatively non-renormalizable. This model
also was shown to go to a trivial model \cite{ko_ko_94,zi_89}.

There are new attempts to make sense of these theories either as
an effective model at low energies, which will give valuable
information in QCD \cite{mu_87, mi_93}, for example for the
studies of hadron mass generation through spontaneous symmetry
breaking. Another attempt is gauging the model and investigating
whether the new coupling gives rise to a nontrivial theory
\cite{ba_le_lo_86,le_lo_ba_86,re_00,re_hepth_99}. These examples
show that  the search for non-trivial models using only fermions
may be an interesting endeavor.

Another attempt in writing a model using only fermions came with
the work of G\" ursey, \cite{gu_56}.  Here a non-polynomial but
conformal invariant Lagrangian was written to describe
self-interacting fermions with the intention of remedying some of
the problems of the Heisenberg model \cite{he_54}.  To be able to
write a conformally invariant lagrangian, G\"{u}rsey had to use a
non-polynomial form. Kortel found solutions to this conformal
invariant theory \cite{ko_56} which were shown to be instantons
and merons much later \cite{ak_82}.

One of us, with collaborators, tried to make quantum sense of this
model a while ago
\cite{ak_ar_du_ho_ka_pa_82-34,ak_ar_du_ho_ka_pa_82-41,ak_ar_ho_pa_83},
finding that even if these attempts were justified, this model
went to a trivial model as the cut-off is removed. Several
processes were calculated \cite{ar_ho_83} involving incoming and
outgoing spinors which gave exactly the naive quark model results,
missing the observed logarithmic behavior predicted by  QCD
calculations.

We tried to give a new interpretation of our old work in
\cite{ho_lu_06}. In that work we saw that the polynomial form of
the original model really did not correspond to the original G\"
ursey model in the exact sense. The two versions  obey different
symmetries. This was shown explicitly in reference
\cite{ho_lu_06}.   We went to higher orders in the calculation,
beyond one loop for scattering processes.  By using the
Dyson-Schwinger and Bethe-Salpeter equations we could calculate
higher order  processes. We saw that while the non-trivial
scattering of the fundamental fields was not allowed, bound states
could scatter from each other with non-trivial amplitudes.

The essential point in our analysis was the fact that, being
proportional to ${{\epsilon}\over{p^2}}$, the composite scalar
field propagator cancelled many of the potential infinities that
arise while calculating loop integrals. As a result of this
cancellation, only  composite fields participate in  physical
processes such as scattering and particle production.  The
scattering and production of  elementary spinor fields were not
allowed. This phenomena was an example of treating the bound
states, instead of the principal fields, as physical entities.

A further point will be to  couple an elementary vector field to
the model described in reference \cite{ho_lu_06}, in line with the
process studied for the Nambu-Jona-Lasinio model
\cite{ba_le_lo_86,le_lo_ba_86}. Coupling the same elementary field
to the model described in reference \cite{ho_ta_hepth_06} will be
similar, giving a model with two vector fields, one composite, the
other one elementary. Our final goal is to investigate if we get a
non-trivial theory when we couple a Yang-Mills system with color
and flavor degrees of freedom, like it is done in
\cite{re_00,re_hepth_99}. Here we study the abelian case, as an
initial step.

In this note we summarize the changes in our results when this
elementary vector field is coupled to the model described in
reference \cite{ho_lu_06}. We outline the model as is given in
Refs. \cite{ak_ar_du_ho_ka_pa_82-34} and \cite{ho_lu_06} in next
section and give our new results in subsequent sections. The main
conclusion is that our original model, in which only the
composites take part in physical processes like scattering or
particle production, is reduced to a gauged--Higgs-Yukawa model,
where both the composites and the fundamental spinor and vector
fields participate in all the processes.
\\
\\
\section{The Model}

Our initial model is given by the Lagrangian

\begin{equation}
L = {i\overline{\psi}} \partial \!\!\!/ \psi + g {\overline{\psi}}
\psi \phi +\xi ( g{\overline {\psi}} \psi -a\phi^{3} ).\label{cl}
\end{equation}

Here the only terms with kinetic part are the spinors. $\xi$ is a
Lagrange multiplier field, $\phi $ is a scalar field with no
kinetic part, $g$ and $a$ are coupling constants. This expression
contains two constraint equations, obtained from writing the
Euler-Lagrange equations for the $\xi$ and $\phi$ fields.  Hence,
it should be quantized using the Dirac constraint analysis as
performed in reference \cite{ho_lu_06}.

The Lagrangian given above is just an attempt in writing the
original G\" ursey Lagrangian

\begin{equation}
 L={i\overline{\psi}} \partial \!\!\!/ \psi + g' ({\overline{\psi}} \psi)^{4/3}
 ,\label{gl}
\end{equation}

in a polynomial form.

We see that the $\gamma^{5} $ invariance of the original
G\"{u}rsey Lagrangian is retained in the form written  in equation
(\ref{cl}).  This discrete symmetry prevents $\psi$ from acquiring
a finite mass in higher orders. We also see that the two models
given by lagrangians in equations (\ref{cl}) and (\ref{gl}) are
not equivalent since the former  does not obey one extra symmetry
obeyed by the latter one. This was carefully studied in reference
\cite{ho_lu_06}. We, therefore, take the first model as a model
which only approximates the original G\"{u}rsey model, without
claiming equivalence and study only that  model in this work.

To quantize the latter system consistently we proceed via the path
integral method. This procedure was carried out in reference
\cite{ho_lu_06}.  At the end of these calculations we found out
that we can write the constrained lagrangian given in equation
(\ref{cl}) as

\begin{equation}
L'' = {i\overline{\psi}}[\partial \!\!\!/  -ig
\Phi]\psi-{{a}\over{16}}(\Phi^{4}+2\Phi^{3}\Xi-2\Phi\Xi^{3}-
\Xi^{4})+{i\over{4}}c^*(\Phi^{2}+2\Phi\Xi+\Xi^{2}) c,
\end{equation}

where the effective lagrangian  is expressed in terms of scalar
fields $\Phi$, and $\Xi$, ghost fields  $c $, $c^*$ and spinor
fields only.

The fermion propagator is the usual Dirac propagator in lowest
order, as can be seen from the Lagrangian.  After integrating over
the fermion fields in the path integral, we obtain the effective
action. The second derivative of the effective action with respect
to the $\Phi$ field gives us the induced inverse propagator for
the $\Phi$ field, with the infinite part given as

\begin{equation} \mbox{Inf} \left[ {{ig^2}\over{ (2\pi)^4}} \mbox{Tr} \int {{d^4
p}\over {p\!\!\!/(p\!\!\!/+q\!\!\!/)}}\right]=
 {{g^2  q^2}\over {8\pi^{2} \epsilon}}.
\end{equation}

Here dimensional regularization is used for the momentum integral
and $\epsilon = 4-n$.  We see that the $\Phi$ field propagates as
a massless field.

When we study the propagators for the other fields, we see that no
linear or quadratic term in $\Xi$ exists, so the one loop
contribution to the $\Xi$ propagator is absent. Similarly the
mixed derivatives of the effective action with respect to $\Xi$
and $\Phi$  are zero at one loop, so no mixing between these two
fields occurs. We can also set the propagators of the ghost fields
to zero, since they give no contribution in the one loop
approximation.  The higher loop contributions are absent  for
these fields.

In reference \cite{ho_lu_06} we also studied the contributions to
the fermion propagator at higher orders and we found, by studying
the Dyson-Schwinger equations for the two point function, that
there were no new contributions.  We had at least one phase where
the mass of the spinor field was zero.

In reference \cite{ho_ta_hepth_06} we studied a similar model
where the composite vector field replaced the composite scalar
field, with similar results.

\section{New Results and Higher Orders}

Here we couple an elementary vector field to the model described
in reference \cite{ho_lu_06}, in a minimal way, with a new
coupling constant $e$,  acting in accordance of the work in
references
\cite{ba_le_lo_86,le_lo_ba_86,re_00,ko_ta_ya_93,ku_te_99}. The new
lagrangian is given as

\begin{eqnarray} L' = {i\overline{\psi}}[ \partial \!\!\!/  -i g  \Phi] \psi -
{{a }\over{4}} (\Phi^4+2\Phi^3 \Xi-2\Xi^3
\Phi-\Xi^4)\nonu \\
+{{i}\over{4}} c^*(\Xi^2 + 2\Phi \Xi +\Phi^2) c-{{1}\over{4}}
F_{\mu \nu} F^{\mu \nu} -{\overline{\psi}}e A\!\!\!/\psi .
\end{eqnarray}

Here $A^{\mu}$ is the elementary vector field and $ F^{\mu \nu}$
is defined from  $A^{\mu}$ in the usual way. We take the vector
field propagator in the Feynman gauge in our explicit
calculations. This lagrangian reduces to the effective expression
given below, since the $\Xi$ and the ghost fields decouple.

\begin{eqnarray} L' = {i\overline{\psi}}[ \partial \!\!\!/  -i g  \Phi] \psi -
{{a }\over{4}} \Phi^4 -{{1}\over{4}} F_{\mu \nu} F^{\mu \nu}
-{\overline{\psi}}e A\!\!\!/\psi . \end{eqnarray}

In this section we summarize the changes in our results for this
new model.

If our fermion field had a color index $i$ where $i=1...N$, we
could perform an 1/N expansion to justify the use of only ladder
diagrams for higher orders for the scattering processes. Although
in our model the spinor has only one color, we still consider only
ladder diagrams anticipating that one can construct a variation of
the model with N colors.

\subsection{Renormalization Group Analysis}

In the models given in references \cite{ho_lu_06} and
\cite{ho_ta_hepth_06}, we had two coupling constants, $g$ and $a$
in reference \cite{ho_lu_06} and only one, which we rename as
$g'$, in reference \cite{ho_ta_hepth_06}. In the model described
in reference \cite{ho_ta_hepth_06}, there is no need for infinite
coupling constant renormalization, since the spinor box diagram is
finite when the incoming and outgoing particles are vectors
\cite{ka_ne_50,wa_50}. In the model described in reference
\cite{ho_lu_06}, the coupling constant $a$  needs renormalization.
In these models there is no need for infinite renormalization for
$g$ and respectively for $g'$ since the diagrams for the
$<{\overline{\psi}}\psi\phi> $ and $<{\overline{\psi}}\psi A_\mu>$
vertices are finite.

Using the language of renormalization group analysis, the first order for this vertex is
given by

\be\mu\frac{dg_0}{d\mu}=0,  \ee

since the diagram given in Figure 1.a is finite, due to the
presence of  $\epsilon$ in the scalar propagator. Higher order
calculations using the Bethe-Salpeter equation verify that the
right hand side of the equation does not change in higher orders.
This process was studied in reference \cite{ho_lu_06}.

We see that in the original model the only infinite
renormalization is needed for the four $\phi$ vertex; hence the
coupling constant for this process {\it runs}. The first
correction to the tree diagram is the box diagram, shown in Figure
1.b . This diagram has four spinor propagators and gives rise to a
${{1}\over {\epsilon}} $ type divergence. The renormalization
group equation written for this vertex is

\begin{eqnarray}
 16\pi^2\mu\frac{da}{d\mu} &=& -dg_0^4 ,
\end{eqnarray}

Here the right hand side of the equation is equal to a constant,
since $g_0$ does not run. Since we include the four $\phi$ term in
our original lagrangian, we can renormalize the coupling constant
of this vertex to incorporate this divergence.  There are no
higher infinities for this vertex. The two loop diagram contains,
as shown in Figure 1.c,  a $\phi$ propagator which makes this
diagram finite. The three-loop diagram is made out of eight spinor
and two scalar lines, Figure 1.d. At worst we end up with a first
order infinity of the form ${{1}\over{\epsilon}}$ using the
dimensional regularization scheme.  Higher order ladder diagrams
give at worst the same type of divergence. This divergence for the
four scalar vertex  can be renormalized using standart means.

\begin{figure}[h]
\begin{center}
$\begin{array}{c@{\hspace{1cm}}c@{\hspace{5mm}}c@{\hspace{5mm}}c}
 \multicolumn{1}{l}{\mbox{\bf }}&
 \multicolumn{1}{l}{\mbox{\bf }}&
 \multicolumn{1}{l}{\mbox{\bf }}&
 \multicolumn{1}{l}{\mbox{\bf }}\\
 [-0.53cm]
 \epsfxsize=16mm \epsffile{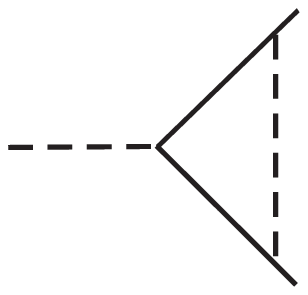} &
 \epsfxsize=25mm \epsffile{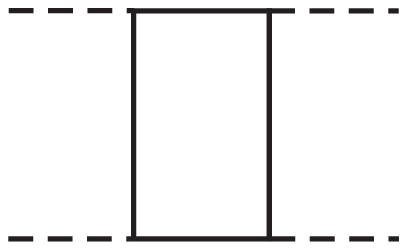} &
 \epsfxsize=25mm \epsffile{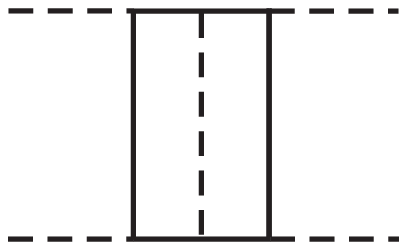} &
 \epsfxsize=40mm \epsffile{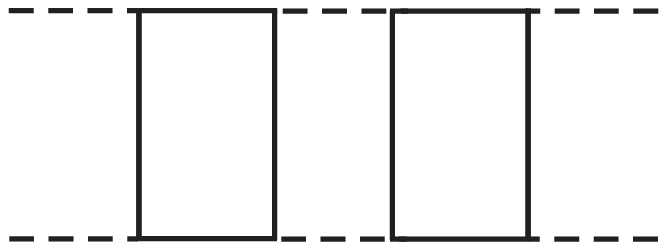} \\
 [0.4cm]
 \mbox{\bf (a)} &
 \mbox{\bf (b)} &
 \mbox{\bf (c)} &
 \mbox{\bf (d)}
 \end{array}$
\end{center}
\caption{The diagrams related to the initial model . Here dotted
lines represent the scalar, solid lines the spinor particles. }
\label{fig475}
\end{figure}

In the new model, where an elementary vector field is added to the
model described in reference \cite{ho_lu_06}, we add a new
coupling constant $e$ which describes the coupling of the vector
field to the spinors. Here  all three coupling constants are
renormalized.

We can write the three first order renormalization group equations
for these three coupling constants similar to the analysis in
\cite{ha_ki_ku_na_94}.

\begin{eqnarray}
 16\pi^2\mu\frac{de}{d\mu} &=& be^3, \\
 16\pi^2\mu\frac{dg}{d\mu} &=& -cge^2,\\
 16\pi^2\mu\frac{da}{d\mu} &=& -dg^4 ,
\end{eqnarray}

where $b$, $c$, $d$ are numerical constants. These values are
given as $b=2$, $c=4$, $d=4$. These processes are illustrated in
diagrams shown in Figure 2. below.

\begin{figure}[h]
\begin{center}
$\begin{array}{c@{\hspace{2cm}}c@{\hspace{2cm}}c}
 \multicolumn{1}{l}{\mbox{\bf }}&
 \multicolumn{1}{l}{\mbox{\bf }}&
 \multicolumn{1}{l}{\mbox{\bf }}\\
 [-0.53cm]
 \epsfxsize=17mm \epsffile{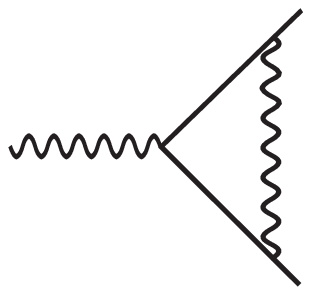} &
 \epsfxsize=17mm \epsffile{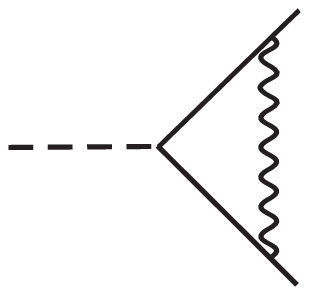} &
 \epsfxsize=25mm \epsffile{g4} \\
 [0.4cm]
 \mbox{\bf (a)} &
 \mbox{\bf (b)} &
 \mbox{\bf (c)}
 \end{array}$
\end{center}
\caption{The three coupling constant corrections in one loop. Here
vector particles are represented by the wiggly lines additional to
the former ones shown in Fig. 1. } \label{fig456}
\end{figure}

Our equations differ from those in reference
\cite{ha_ki_ku_na_94}, since the interaction of the composite
scalar field with the spinors does not result in infinite terms
due to the presence of the factor $\epsilon$ in the scalar
propagator.  Here $\epsilon$, the parameter in the dimensional
regularization scheme, is inversely proportional to $\ln
{{\Lambda}\over {\Lambda_0}} $ where $\Lambda$ is the cut-off
parameter. These equations have the immediate solutions

\begin{eqnarray}
 e^2 &=& {{e^{2}_{0} }\over {A}}, \\
 g   &=&g_0 A^{c/2b}, \\
 a   &=& a_0 + \frac{dg_0^4}{2(2c+b)e_0^2} A^{\frac{2c}{b}+1} \label{ega1} \end{eqnarray}

where

\begin{eqnarray} A= 1-\frac{2b e^2_0}{16\pi^2}\ln{{\mu}\over{\mu_0}}. \end{eqnarray}

If we use diagrammatical analysis, we see that only the
spinor-vector field coupling gives infinite contribution to the
first two equations. The third equation diverges because of the
contribution of the box diagram, which is infinite even in the
absence of the vector field. For this coupling constant, at one
loop level, there is no difference from its behavior in the
original model.

In the original model we need an infinite renormalization for only
one of the coupling constants, the one with the four scalar field.
Further renormalization may be necessary at each higher loop, like
any other renormalizable model. The difference between our
original  model and other renormalizable models lies in the fact
that, although this model is a renormalizable one using naive
dimensional counting arguments, we have only one set of diagrams
which is divergent. We need to renormalize only one of the
coupling constants by an infinite amount. This set of diagrams,
corresponding to the scattering of two bound states to two bound
states, has the same type of divergence, i.e.
${{1}\over{\epsilon}}$ in the dimensional regularization scheme
for all odd number of loops. The contributions from even number of
diagrams are finite, hence require no infinite renormalization.

When additional the vector particle contributions are added, this
expression is modified. The process where two scalar particles
goes to two scalar particles gets further infinite contributions
from the box type diagrams with vector field insertions, where one
part of the diagram is connected to the non-adjacent  part with a
vector field as shown in the Figure 3.a. All these diagrams go as
${{1}\over {\epsilon}}$ where $\epsilon$ is the parameter in
dimensional regularization scheme. There are no higher divergences
for this process. Note that mixed scalar and vector insertions do
not give additional infinities, since the scalar propagator
reduces the degree of divergence.  Also note that the diagram
where the internal photon is connecting adjacent sides, as shown
Figure 3.b, will be a contribution to the coupling constant
renormalization of one of the vertices. Since this is not a new
contribution, we will not consider it separately.

\begin{figure}[h]
\begin{center}
$\begin{array}{c@{\hspace{2cm}}c}
 \multicolumn{1}{l}{\mbox{\bf }}&
 \multicolumn{1}{l}{\mbox{\bf }}\\
 \epsfxsize=25mm \epsffile{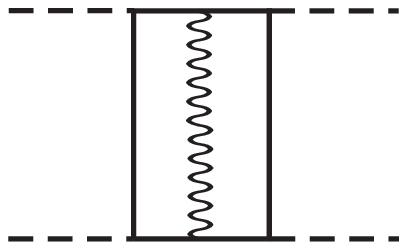} &
 \epsfxsize=25mm \epsffile{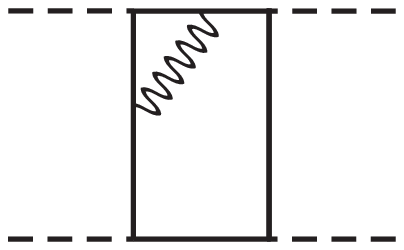} \\
 [0.4cm]
 \mbox{\bf (a)} &
 \mbox{\bf (b)}
 \end{array}$
\end{center}
\caption{(a) The vector particle correction to the fermion box
diagram, (b) the box diagram with one vertex correction.  }
\label{fig2}
\end{figure}

\subsection{Propagators and Vertices}

We have no essential change in the spinor propagator. In reference
\cite{mi_93} Miransky explains how for coupling constant $\alpha$
less than $\pi/3$, there is no mass generation in the quenched
approximation.  Here $\alpha= {{e^2}\over{4\pi}}$. J.C.R. Bloch,
in his Durham thesis, \cite{bl_hepph_02}, explores the range where
this result is valid when the calculation is done without this
approximation. He states that the quenched and the rainbow
approximations, used by Miransky and collaborators, have non
physical features, namely they are not  gauge invariant, making
the calculated value wildly vary depending on the particular gauge
used.  Bloch, himself,   uses the Ball-Chiu vertex,
\cite{ba_ch_80}, instead of the bare one,  where the exact
longitudinal part of the full QED vertex, is uniquely determined
by the Ward-Takahashi identity relating the vertex with the
propagator.  The transverse part of the vertex, however, is still
arbitrary. Bloch then considers a special form of the
Curtis-Pennington vertex \cite{cu_pe_90} in which the transverse
part of the vertex is constructed by requiring the multiplicative
renormalizability of the fermion propagator with additional
assumptions.

Bloch claims that for the different gauges used with this choice,
he gets rather close values for the critical coupling
\cite{ak_bl_gu_pe_re_94}. He also performs numerical calculations
where the approximations are kept to a minimum. The results are
given in the table on pg. 202 of hep-ph/0208074.

Using on the arguments in the Bloch's thesis, also using the
results of his numerical calculations, we conclude that at least
for $\alpha <  0.5 $ we can safely claim that there will be no
mass generation or the assumed $\gamma_5$ symmetry will be not
broken. Since we do not study  heavy ion processes, the numerical
value we have for $\alpha$ will be much smaller than this limit.
Hence,  our results will be valid.  Note that in QCD mass
generation occurs at relatively low energies, where the coupling
constant has already increased.

Miransky \cite{mi_93} also explains how in the Landau gauge we can
take the coefficient of the momentum term  as unity. Using these
arguments we can conclude that there are no additional
contributions to the spinor propagator used in reference
\cite{ho_lu_06}, at least in the Landau gauge.

The photon propagator also will be the similar as the one given in
QED, with only additional ${{1}\over{\epsilon}}$ contributions
from the scalar particle insertions.  The lowest order diagram for
this process is shown in Figure 4.a. The dominant contribution
will be from the vector insertions, which are studied in QED.

\begin{figure}[htb!]
\begin{center}
$\begin{array}{c@{\hspace{2cm}}c}
 \multicolumn{1}{l}{\mbox{\bf }}&
 \multicolumn{1}{l}{\mbox{\bf }}\\
 [-0.53cm]
 \epsfxsize=25mm \epsffile{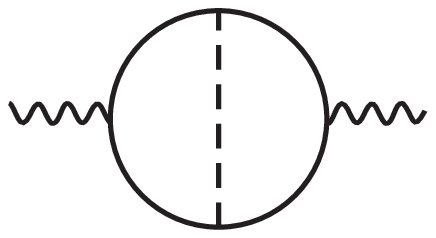} &
 \epsfxsize=25mm \epsffile{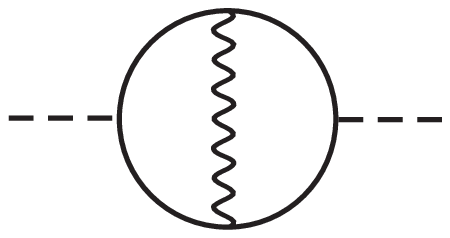} \\
 \end{array}$
\end{center}
\caption{ (a) Scalar contribution to the vector propagator, (b)The
vector particle correction to the scalar propagator}
\end{figure}

The additional contribution to the scalar propagator can be
calculated using diagrammatical analysis.  If we take only the
planar diagrams which connect two different spinor lines, as shown
in Figure 4.b, the scalar field contributions are only of order
${{1}\over {\epsilon}}$, the same as the one loop initial
contribution. Higher order divergences come from the vector field
insertions.

The higher order planar insertions will be the dominant ones if we
allow $N_f$ flavors for the fermions, where $N_f$ is large,  and
perform an ${{1}\over{N_f}}$ expansion. We will assume that the
same approximation can be done in our case too.  The diagrams
where there are $n-1$ nonadjacent and planar vector field
contributions, go as $({{-D}\over {\epsilon}})^n$, where $D=
{{-4e^2}\over {(4\pi)^2}}$ is a numerical constant. Naively the
planar vector field contributions can  be summed up as a geometric
series \cite{ka_va_hepth_06}. The same result is true also for the
planar vertex corrections as in Figure 5.a.

The vector- spinor- antispinor vertex do not get infinite
contributions from our composite scalar particle.  A typical
diagram is given in Figure 5.b. Here the infinities coming from
the integrations are cancelled by the $\epsilon$ factors in the
scalar propagators. That vertex, for the purely electromagnetic
case, Figure 5.c, is vastly studied in the literature
\cite{ba_ch_80,cu_pe_90}.

\begin{figure}[htbp]
\begin{center}
$\begin{array}{c@{\hspace{2cm}}c@{\hspace{2cm}}c}
 \multicolumn{1}{l}{\mbox{\bf }}&
 \multicolumn{1}{l}{\mbox{\bf }}&
 \multicolumn{1}{l}{\mbox{\bf }}\\
 [-0.53cm]
 \epsfxsize=20mm \epsffile{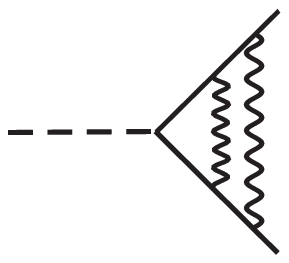} &
 \epsfxsize=20mm \epsffile{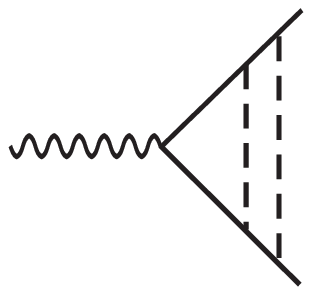} &
 \epsfxsize=20mm  \epsffile{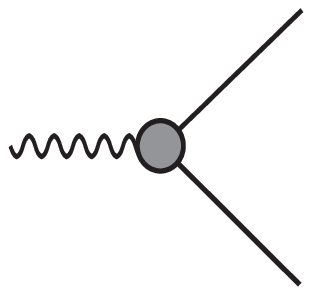} \\
 [0.4cm]
 \mbox{\bf (a)} &
 \mbox{\bf (b)} &
 \mbox{\bf (c)}
 \end{array}$
\end{center}
\caption{ (a) The vector particle correction in higher orders, (b)
The scalar particle insertions to the vector- spinor- antispinor
vertex, (c) The vector- spinor-antispinor vertex } \label{fig3}
\end{figure}

\subsection{Scattering and Production Processes}

The process where two composite scalars scattering from each other
was studied above.  The scattering of two scalars producing four,
or to any higher even number of scalars is finite, as expected to
have a renormalizable model. The process where two scalars create
an odd number of scalars is forbidden by the $\gamma^5$ invariance
of the theory, hence two scalar $\phi$ particles can only go to an
even number of scalar particles. This assertion is easily checked
by diagrammatic analysis.

We also note that in the original model the four spinor kernel was
of order $\epsilon $. The lowest order diagram, shown in Figure
6.a, vanishes as $\epsilon$ due to the presence of the scalar
propagator. In higher orders this expression can be written in the
quenched ladder approximation \cite{mi_93}, where the kernel is
seperated into a scalar propagator with two spinor legs joining
the proper kernel. If the proper kernel is of order $\epsilon$,
the loop involving two spinors and a scalar propagator can be at
most finite that makes the whole diagram in first order in
$\epsilon$. This fact also shows that there is no nontrivial
spinor-spinor scattering.

As a result of this analysis, in the ungauged version,  we end up
with a model where there is no scattering of the fundamental
fields, i.e. the spinors, whereas the composite scalar fields can
take part in a scattering process.  The coupling constant for the
scattering of the composite particles runs, whereas the coupling
constant for the spinor-scalar interaction does not run. The
processes giving this conclusion are carefully studied in
reference \cite{ho_lu_06}.

This result changes drastically when the gauged model is studied
instead of the original one. This process, which is prohibited in
the previous model, \cite{ho_lu_06}, now is possible due to the
presence of the vector field channel. In lowest order this process
goes through the tree diagram given in Figure 6.b.

The process is finite though, since at the next higher order the
QED box diagram with two spinors and two vector particles, Figure
6.c,  is ultra violet finite from dimensional analysis, and is
calculated in reference \cite{po_ru_02}. Higher orders do not give
new type of ultra violet divergences.

We also allow spinor production from the scattering of scalar
particles, since now we can use vector particles as
intermediaries, Figure 6.d.

\begin{figure}[htbp]
\begin{center}
$\begin{array}{c@{\hspace{1cm}}c@{\hspace{1cm}}c@{\hspace{1cm}}c}
 \multicolumn{1}{l}{\mbox{\bf }}&
 \multicolumn{1}{l}{\mbox{\bf }}&
 \multicolumn{1}{l}{\mbox{\bf }}&
 \multicolumn{1}{l}{\mbox{\bf }}\\
 [-0.53cm]
 \epsfxsize=26mm \epsffile{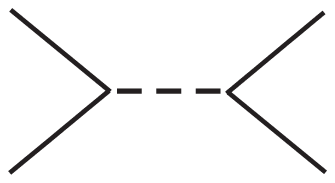} &
 \epsfxsize=26mm \epsffile{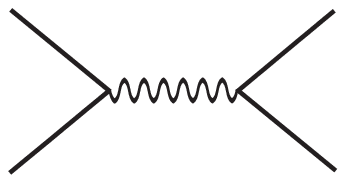} &
 \epsfxsize=17mm \epsffile{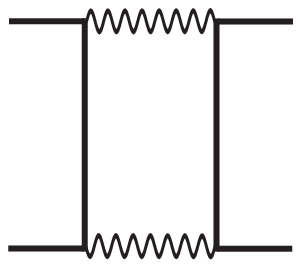} &
 \epsfxsize=25mm \epsffile{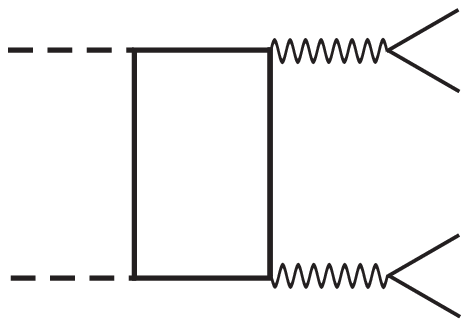} \\
 [0.4cm]
 \mbox{\bf (a)} &
 \mbox{\bf (b)} &
 \mbox{\bf (c)} &
 \mbox{\bf (d)}
\end{array}$
\end{center}
\caption{(a) Two fermion scattering through the scalar particle
channel, (b) Two fermion scattering through the vector particle
channel, (c) Higher order diagram for two spinor scattering, (d)
Spinor production from scattering of scalars} \label{fig4}
\end{figure}

\section{Conclusion}

In this note we discussed  the differences  between the  new
model, introduced in this paper,  and the model studied in
reference \cite{ho_lu_06}. We found out that many of the features
of the original model are not true anymore.  As far as
renormalizations are concerned, we have essentially QED, with
corrections coming from the scalar part mimicking the Yukawa
interactions with the $\Phi^4$ term added.  We end up with the
gauge-Higgs-Yukawa system, although our starting point is gauging
a constrained model.

We  also have scattering processes where two scalar particles go
to an even number of scalar particles, or scattering of spinor
particles from each other. In the one loop approximation all these
diagrams give finite results, like the case in the standard Yukawa
coupling model. We also have creation of spinor particles from the
interaction of scalars, as well as scattering of spinors with each
other, and all the other processes in the gauge-Higgs-Yukawa
system.

If we consider the model described in reference
\cite{ho_ta_hepth_06}, we see that the same differences prevail.
The main results are the same. The only difference from the scalar
model is the finiteness of the spinor box diagram with incoming
and outgoing vector particles, \cite{ka_ne_50,wa_50},  both in the
new model and the one in reference \cite{ho_ta_hepth_06}.

\vspace{5mm}{\bf{Acknowledgement}}: The work of M.H. is also
supported by TUBA, the Academy of Sciences of Turkey. This work is
also supported by TUBITAK, the Scientific and Technological
Council of Turkey.

\end{document}